\documentclass{article}
\usepackage{nicefrac}
\usepackage{amsfonts}
\usepackage{amssymb}
\usepackage{graphicx}
\usepackage{appendix}
\usepackage{cite}
\usepackage[intlimits]{amsmath}
\usepackage[intlimits]{amsmath}
\begin{document}

\title{Thermodynamics of the quantum $su(1,1)$ Landau-Lifshitz model.}
\author{A. Melikyan\thanks{amelik@fis.unb.br} and A. Pinzul\thanks{apinzul@unb.br} \\
\\
\emph{Universidade de Bras\'{\i}lia}\\
\emph{Instituto de F\'{\i}sica}\\
\emph{70910-900, Bras\'{\i}lia, DF, Brasil}\\
\emph{and}\\
\emph{{International Center of Condensed Matter Physics} }\\
\emph{C.P. 04667, Brasilia, DF, Brazil} \\
}
\date{}
\maketitle

\begin{abstract}
We present thermodynamics of the quantum $su(1,1)$ Landau-Lifshitz model, following
our earlier exposition [J. Math. Phys. 50, 103518 (2009)] of the quantum
integrability of the theory, which is based on construction of self-adjoint
extensions, leading to a regularized quantum Hamiltonian for an
arbitrary $n $-particle sector. Starting from general discontinuity properties of the
functions used to construct the self-adjoint extensions, we derive the
thermodynamic Bethe Ansatz equations. We show that due to non-symmetric and singular kernel, the self-consistency implies that only negative chemical potential values are allowed, which leads to the conclusion that, unlike its $su(2)$ counterpart, the $su(1,1)$ $LL$ theory at $T=0$ has no instabilities.
\end{abstract}

\newpage

\section{Introduction}

There has been a renewed interest in the Landau-Lifshitz ($LL$) model in
recent years, mainly due to its integrability, and, thus, its role as a
toy-model to check the gauge/string duality as both having underlying
integrable structures (see for recent reviews \cite{spec.issue:2009,Arutyunov:2009ga}). A large amount of literature has been devoted to
this subject, mostly concerning its classical integrability aspects. At this point, there
is relatively complete understanding of the classical integrable properties
of the $LL$ model \cite{Faddeev:1987ph,kosevich1990ms,mikeska2004odm}. The quantum integrability, on the other hand, has been
shown to be surprisingly complex and non-trivial.\footnote{
See an extensive list of literature in \cite{Melikyan:2008ab}} Firstly, as demonstrated
by Sklyanin \cite{Sklyanin:1988s1}, one has to clearly distinguish the $su(1,1)$ and the $su(2)$
models, as they have completely different physical properties. This is due
to construction of the positive-definite metric in the corresponding Hilbert
spaces. So far, only the $su(1,1)$ case has been addressed in literature
from this point of view, as in the $su(2)$ case the construction of the
positive-definite metric seems to be a quite complicated mathematical task \cite{albeverio1983frp}.
The next interesting issue arises in the \emph{anisotropic} $LL$ model, where
the algebra of observables has to be modified for
consistency to be a quadratic one. While natural from the lattice regularization point of view \cite{Sklyanin:1982tf,Sklyanin:1983ig}, it
is not clear how to derive it directly within the framework of the
continuous integrable model. And finally, the $LL$ model is distinctly
different from other known integrable models due to its highly singular
nature in the quantum mechanical description \cite{Korepin:1997bk}. Indeed, it gives rise to $\delta ^{\prime \prime }(x)$ types of interactions which are not mathematically easy to deal with \cite{Albeverio:1988}. As a result, the quantum Hamiltonian was
not possible to obtain until recently.

In our previous paper \cite{Melikyan:2008ab}, we have solved one of the outlined above problems,
and by constructing the self-adjoint extensions with the help of the unusual vector-like
states in Hilbert space, specified by the non-trivial scalar product, we had
explicitly presented the quantum Hamiltonian, and derived the spectrum of the $su(1,1)$ isotropic model, in
complete agreement with all known results. In the current paper we work out
the thermodynamics of the model, relying on our previous construction. We
re-derive the Bethe Ansatz equations, find the integral equation
describing the excitation and finally construct the thermodynamics,
following the standard methods. We show that unlike other models, in this case one has to start from a finite temperature case, and only then consider a careful $T \longrightarrow 0$ limit to avoid contradictory results. One interesting consequence of the thermodynamic Bethe Ansatz equations for the $su(1,1)$ $LL$ model is that there is no Fermi point, which means that only the negative values for chemical potential are allowed. We use this feature to argue that the $su(1,1)$ field theory at $T=0$ has no instabilities, unlike the perturbative results \cite{Klose:2006dd} for the $su(2)$ model.

We emphasize that we use $LL$ model as a
testing ground for such singular potentials, as the above mentioned aspects
and difficulties of the $LL$ model can be equally transferred to other interesting
models, e.g. the fermionic $AAF$ model \cite{Alday:2005jm}, quantum mechanical description of which has the
same highly singular $\delta ^{\prime \prime }(x)$ behavior. Our goal here
is to learn to deal with quantum integrability directly in the continuous
case, without resorting to the lattice regularization schemes, which can be
a daunting task by itself. We also consider only the $su(1,1)$ isotropic $LL$
model, leaving the essential complications of the $su(2)$ and anisotropic
cases to a future work.

Our paper is organized as follows. In section \textbf{2}, we give a short
description of the Landau-Lifshitz model, and give only essential details of
our method of constructing the self-adjoint extensions and give the quantum
Hamiltonian, as well as the exact spectrum. On this basis, in section
\textbf{3}, we use the discontinuity properties of the functions used to
construct the Hilbert space, re-derive in this way the Bethe equations, and
following this initial setup work out the thermodynamics. Finally, in conclusion,
we remark that the results of the current paper open several interesting questions, and we outline possible directions.
\clearpage
\section{The $LL$ model, quantum Hamiltonian and self-adjoint extensions.}

The classical Hamiltonian of the full anisotropic Landau-Lifshitz model,\footnote{We use the notations of \cite{Sklyanin:1988s1}} has the form:

\begin{equation}
H=\frac{\epsilon }{2}\int \left[ -\left( \partial _{x}\mathbf{S},\partial
_{x}\mathbf{S}\right) +4\gamma ^{2}\left( \left( S^{3}\right) ^{2}-1\right)
\right]  \label{Init.Ham.}
\end{equation}
Here the scalar product $(\mathbf{S,S})\equiv \left( S^{3}\right)
^{2}-\epsilon \left( S^{1}\right) ^{2}-\epsilon \left( S^{2}\right) ^{2}=1$
for the vector $\mathbf{S}=(S^{1},S^{2},S^{3})$ depends on the parameter $
\epsilon =\pm 1,$ which gives rise to two different models corresponding to
the $su(1,1)$ and $su(2)$ cases. In what follows we will consider only the $
su(1,1)$ ($\epsilon = 1$) and isotropic ($\gamma =0$) case, as there are, as
we noted above, principal difficulties with $su(2)$ or anisotropic cases in
the quantum theory. The classical dynamics is governed by the Poisson
structure:
\begin{align}
\{S^{3}(x),S^{\pm }(y)\}& =\pm iS^{\pm }(x)\delta (x-y)  \notag \\
&  \label{P.B.} \\
\{S^{-}(x),S^{+}(y)\}& =2i S^{3}(x)\delta (x-y)  \notag
\end{align}
which is replaced, in the quantum theory, by the commutation relations for
the $S$-operators:\footnote{
This is true only for $\gamma =0.$ In general, the algebra (\ref{Com.rel.})
should be modified to a quadratic \emph{Sklyanin} algebra \cite{Sklyanin:1982tf}.}
\begin{align}
\left[ S^{3}(x),S^{\pm }(y)\right] & =\pm S^{\pm }(x)\delta (x-y)  \notag \\
&  \label{Com.rel.} \\
\left[ S^{-}(x),S^{+}(y)\right] & =2 S^{3}(x)\delta (x-y)  \notag
\end{align}

The representations of (\ref{Com.rel.}), in the ferromagnetic vacuum $
S^{3}(x)|0\rangle =|0\rangle ;$ $S^{-}(x)|0\rangle =0$, can be constructed
using the vectors of the form
\begin{equation}
|f_{n}\rangle =\int
dx_{1}...dx_{n}f_{n}(x_{1...}x_{n})S^{+}(x_{1})...S^{+}(x_{n})|0\rangle
\label{representation}
\end{equation}
where $f_{n}(x_{1...}x_{n})$ are continuous and decreasing sufficiently fast
functions for the integral (\ref{representation}) to be well defined. It can
be shown that only the $su(1,1)$ case guarantees a positive definite scalar
product $\langle g_{n}|f_{n}\rangle $.

The problem one generally faces in continuous integrable models is that the
quantum trace identities, defined via the standard procedure of first
finding the transfer matrix $T(u)$ and then taking the trace, lead, unlike
the classical case, to polynomials which contain operator product at the
same point, and therefore are not well-defined. This issue, however, was
neglected for a long time in the majority of the classical integrable
models, as for example the Thirring model endowed with the $\delta (x)$ type
of potential, by using various (mathematically not well defined) procedures
of taking regularizations such as $\epsilon (x)\delta (x)$ $=0.$ In the $LL$
model, as well as in the fermionic $AAF$ model \cite{Alday:2005jm}, this cannot be done anymore,
and formal manipulations with singular functions lead to meaningless
expressions. As a result of these principal difficulties, the quantum
Hamiltonian for the $LL$ model, as well as other charges, were not possible
to extract from the trace identities until recently.

In our previous paper we partially solved this problem, by finding the
regularized continuous quantum Hamiltonian\footnote{
We will present a method to construct the regularized charges in all orders
in the upcoming paper \cite{melpinzweb:2010}} and giving the
complete description of the resulting Hilbert space. We refer the reader to
the publication \cite{Melikyan:2008ab} for complete details, and only present here the
necessary formulas and essential results. There are essentially two steps
involved in our construction. First, we regularize the continuous
Hamiltonian by the split-point method so that the quantum Hamiltonian has
the form:
\begin{equation}
H_{Q}=\underset{\varepsilon \rightarrow 0}{\lim }H_{\varepsilon }
\label{Quant.Ham.}
\end{equation}
where
\begin{equation}
H_{\varepsilon }=\frac{1}{2}\int dudvF_{\varepsilon }(u,v)\left[ -\partial
_{u}S^{3}\partial _{v}S^{3}+\partial _{u}S^{+}\partial _{v}S^{-}+\partial
_{u}\partial _{v}\left( S^{3}(u)\delta (u-v)\right) -\partial _{u}\partial
_{v}\delta (u-v)\right]  \label{Quant.Ham2a.}
\end{equation}
Here the function $F(u,v)$ is any sufficiently rapidly decreasing smooth
function, which makes the integral well-defined, and the dependence on an
arbitrary parameter (or set of parameters) $\varepsilon $ is not essential
as long as we require that
\begin{equation}
\underset{\varepsilon \rightarrow 0}{\lim }F_{\varepsilon }(u,v)=\delta (u-v)
\label{F(u,v)}
\end{equation}

This condition simply means that the quantum Hamiltonian $H_{Q}$ is local.
It can be shown that the action of the Hamiltonian $H_{Q}$ on an arbitrary $n
$-particle state
\begin{equation}
|f_{n}\rangle =\int d^{n}\vec{x}\ f(\vec{x})\ S^{+}(x_{1})\cdots
S^{+}(x_{n})|0\rangle  \label{npstate}
\end{equation}
has the form
\begin{eqnarray}
H_{Q}|f_{n}\rangle &=&-\int d\overrightarrow{x}\left( \Delta f(
\overrightarrow{x})\right) \overset{n}{\prod_{i=1}}S^{+}(x_{i})|0\rangle
\label{Hq} \\
&&+\sum_{i>j}\int \prod_{k\neq j}dx_{k}\left( \left( \partial _{j}f(\overrightarrow{x})-\partial _{i}f(\overrightarrow{x})\right)
|_{x_{j}=x_{i}-\epsilon }^{x_{j}=x_{i}+\epsilon }+\partial _{i}\partial
_{j}f(\overrightarrow{x})|_{x_{i}=x_{j}}\right) \overset{n}{\prod_{i=1}}
S^{+}(x_{i})|_{x_{i}=x_{j}}|0\rangle  \notag
\end{eqnarray}
(here the Laplacian $\Delta =\sum_{i=1}^{N}\partial _{i}^{2})$ leading to
the following matching conditions:
\begin{equation}
-\left( \partial _{j}f(\vec{x})-\partial _{i}f(\vec{x})\right)
_{x_{j}=x_{i}-\epsilon }^{x_{j}=x_{i}+\epsilon }=\partial _{j}\partial _{i}f(
\vec{x})|_{x_{i}=x_{j}}\ ,\ \ \forall i>j  \label{matching.cond}
\end{equation}
This was shown to be equivalent to self-adjointness of some Hamiltonian $
\widehat{H}$ (for the explicit expression for $\widehat{H}$ see \cite{Melikyan:2008ab},
here we will not need it) acting on the Hilbert space generated by vectors
of the type:
\begin{equation}
\Psi =\left(
\begin{array}{c}
f_{1}(x_{1}) \\
f_{2}(x_{1},x_{2}) \\
...
\end{array}
\right)  \label{vec1}
\end{equation}
with the scalar product, defined, e.g. \ for $n=2$ case as
\begin{equation}
\langle \Phi |\Psi \rangle =\frac{1}{2}\int_{-\infty }^{\infty }g_{1}^{\ast
}(x)f_{1}(x)\ dx+\iint_{x\neq y}g_{2}^{\ast }(x,y)f_{2}(x,y)\ dxdy
\label{scal.pro}
\end{equation}
which was shown to reproduce beautifully Sklyanin's earlier attempts to deal
with the ill-defined object in the $LL$ model. The matching conditions (\ref
{matching.cond}) were also shown to correspond to the the $n$-particle $S$
-matrix factorization property
\begin{equation}
S_{n}(p_{1},...,p_{n)}=\underset{i\neq j}{\prod }S_{2}(p_{i},p_{j})
\label{fact1}
\end{equation}
where the two-particle scattering S-matrix for the $su(1,1)$ $LL$ model can be
derived from (\ref{npstate}) for $n=2$ and has the form:
\begin{equation}
S_{2}=\frac{2(p_{1}-p_{2})+ip_{1}p_{2}}{2(p_{1}-p_{2})-ip_{1}p_{2}}
\label{fact2}
\end{equation}
Let us also note here, that in terms of the cluster fields $\Psi
_{1}^{+}(x_{1})$ and $\Psi _{2}^{+}(x_{1})$ (see for details \cite{Sklyanin:1988s1}) the state $\Psi $
(\ref{vec1}) for $n=2$ can be written in the form
\begin{align}
\Psi _{n=2}& =-\frac{p_1 p_2}{2}\left[ \int dxe^{(p_{1}+p_{2})x}\Psi
_{2}^{+}(x)\right. \\
& \left. +\int_{x_{x}>x_{2}}dx_{1}dx_{2}\left(
c(p_{1,}p_{2})e^{i(p_{1}x_{1}+p_{2}x_{2})}+\overline{c}
(p_{1,}p_{2})e^{i(p_{1}x_{2}+p_{2}x_{1})}\right) \Psi _{1}^{+}(x_{1})\Psi
_{1}^{+}(x_{2})\right] |0\rangle  \notag
\end{align}
where $c(p_{1,}p_{2})=\frac{2(p_{1}-p_{2})+ip_{1}p_{2}}{2(p_{1}-p_{2})}.$
Thus, the $S$-matrix has the meaning of scattering of the quanta
corresponding to the $\Psi _{1}^{+}(x)$ cluster field. Note also, that the
first term is usually absent in other models.

To sum up, our method, being mathematically strict, gives clear systematic
description and understanding of the properties of integrable systems
endowed with singular potential.

\section{Thermodynamics}

Having obtained the spectrum and the $S$-matrix, we proceed to
thermodynamics of the model, following the standard methods (see, for
example, \cite{Korepin:1997bk,Essler:2005fk,Yang:1968rm}). One can easily repeat the analysis of \cite{Korepin:1997bk}, and we
present only the final results and emphasize the differences with the non-linear Schr\"{o}dinger
case. Note, that while the analysis of the basic thermodynamic properties is
quite similar to other models, there are several interesting features that require a careful analysis. In particular, we show below that one cannot start directly from the $T=0$ case, as it leads to contradictory results. Instead, due to singular behavior of the kernel, one has to consider the general $T\neq 0$ case, and then take the $T \longrightarrow 0$ limit. Note also, that essential differences will arise when
considering the correlation functions at finite temperature, as it would
involve the properties of the correct Hilbert space (\ref{vec1}) with the
scalar product (\ref{scal.pro}).

Using the results of the previous section, we obtain the Bethe equations for
the $su(1,1)$ $LL$ model in the system of the length $L:$
\begin{equation}
e^{ip_{i}L}=\underset{i\neq j}{\overset{N}{\prod }}\frac{
2(p_{i}-p_{j})+ip_{i}p_{j}}{2(p_{i}-p_{j})-ip_{i}p_{j}}  \label{be}
\end{equation}
Note that the $S$-matrix for the $su(1,1)$ that we consider here is
different from the one obtained in $su(2)$ by a factor ``$-2$'' in front of $(p_{i}-p_{j})$ \cite{Klose:2006dd}.\footnote{The perturbative calculations for the $su(2)$ model lead to the S-matrix: $S=\underset{i\neq j}{\overset{N}{\prod }}\frac{
(p_{i}-p_{j})-ip_{i}p_{j}}{(p_{i}-p_{j})+ip_{i}p_{j}}$.} (This is not
surprising, considering these are two different models.) While the absolute value of this factor is not of a great importance, the sign plays a big role, essentially leading to different physics: it
immediately follows from (\ref{be}) that unlike the $su(2)$ case, in the $
su(1,1)$ theory the momenta $p$ are always real and as a consequence the
theory does not have bound states.\footnote{It can be shown that this result is true in general for the anisotropic $su(1,1)$ $LL$ model
as well.} It is convenient to introduce variables $u\equiv 1/p.$\footnote{It is obvious already from (\ref{be}) that $p_i = 0$ is a special point in momentum space. For the moment, we will consider this space as $\mathbb{R}\backslash\{0\}$, so that the $u$-variables are well-defined, and extend to the full $\mathbb{R}$ later.} This is
because in the general anisotropic case, the Bethe equations have simpler
form in terms of the $u$ variable, which is associated with the spectral
parameter in the inverse scattering method. Thus, (\ref{be}) can be also
written in the equivalent form:
\begin{equation}
\frac{L}{u_{j}}+\underset{k\neq j}{\overset{N}{\sum }}\Theta
(u_{j}-u_{k})=2\pi n_{j}  \label{be2}
\end{equation}
where $\Theta (u)=i\ln (\frac{i/2-u}{i/2+u}),$ and the numbers $
n_{j}=n_{j}^{\prime }+\frac{N-1}{2},$ $n_{j}^{\prime }\in \mathbb{Z}$, $j=1...N.$
Since the kernel is
\begin{equation}
K(u)\equiv \partial _{u}\Theta (u)=-\frac{1}{1/4+u^{2}}  \label{ker}
\end{equation}
we see that the function $\Theta (u)$ is a monotonically decreasing function
of $u.$ This is contrary to the behavior of the closely related (in fact,
equivalent on the classical level) non-linear Schr\"{o}dinger model, for
which the corresponding $\Theta $ function is a monotonically increasing
function of momenta.
Nevertheless, the solutions can be shown to exist and \ unique due to
existence of the convex Yang-Yang type of action:
\begin{equation}
S_{YY}=-L\underset{j}{\sum }\ln u_{j}+2\pi \underset{j}{\sum }n_{j}u_{j}-
\frac{1}{2}\underset{k\neq j}{\sum }\underset{0}{\overset{u_{j}-u_{k}}{\int }
}\Theta (v)dv  \label{JaJa}
\end{equation}
Using (\ref{be2}) and (\ref{ker}) one can obtain the estimate
\begin{equation}
\frac{2\pi (n_{j}-n_{k})}{L}\geq \frac{1}{u_{j}}-\frac{1}{u_{k}}
=p_{j}-p_{i}\geq \frac{2\pi (n_{j}-n_{k})}{L(1+D)}\geq \frac{2\pi }{L(1+D)}
\label{estim}
\end{equation}
This is exactly the same estimate as in the case of the $NLS$ model, which allows us to
consider the thermodynamic limit of the model. To proceed, one introduces
the density of states in complete analogy as
\begin{equation}
\rho (p_{k})=\frac{1}{L(p_{k+1}-p_{k})}  \label{dens}
\end{equation}
and then takes the thermodynamic limit $L\longrightarrow \infty ;$ $
N\longrightarrow \infty ;$ $D\equiv N/L=(const)$. Because the zero temperature case is rigorously obtained as a limit of the $T\ne 0$ system, we will first consider this general situation, i.e., the derivation of the thermodynamic Bethe
Ansatz ($TBA$) equation. The initial steps of the construction are parallel to those in \cite{Korepin:1997bk} for the $NLS$ system, and we only mention one
interesting difference, that in the process of the minimization of the
partition function one has to be more careful, since the effective kernel $
\widetilde{K}(p_{j},p_{k}) \equiv -\frac{1}{(p_j)^2}{K}(p_{j},p_{k}) = \frac{1}{p_{j}^{2}}\frac{1}{
1/4+(1/p_{j}-1/p_{k})^{2}}$ that appears in the equations in our case is not symmetric under the exchange of the momenta $p_{j}$ and
$p_{k}.$ The resulting $TBA$ equation has the form:
\begin{equation}
\varepsilon (p)=p^{2}-h - \frac{T}{2\pi }\underset{-\infty}{\overset{\infty}{
\int }}d\mu \widetilde{K}(\mu , p )\ln (1+e^{-\varepsilon (\mu )/T})
\label{tba}
\end{equation}

At this point we should recall that the point $p=0$ has been excluded so far from our momentum space. Thus, we have to verify that the solution to (\ref{tba}) is continuous at this point. This will lead to a severe constrain on the chemical potential $h$. To demonstrate this, let us note that the naive conclusion that $\varepsilon (p)\longrightarrow -h$ as $p\longrightarrow 0$ (which seems to follow from the fact that $\widetilde{K}(\mu ,
0 )=0$) is not correct. In fact, we have the following behavior of the kernel when $p\longrightarrow \pm 0$:
\begin{equation}
 \lim_{p\rightarrow \pm 0}\widetilde{K}(\mu , p )=2\pi \delta (\mu \mp 0)
\label{kersing}
\end{equation}
Using (\ref{kersing}) in (\ref{tba}), we obtain
\begin{equation}
\varepsilon(\pm 0) = T \ln \left( e^{-{h}/{T}}-1\right)
\label{epsilon}
\end{equation}
First consequence of (\ref{epsilon}) is that if a solution exists, then it is continuous at the point $p=0$. Now we have two distinct cases:
\begin{itemize}
\item[i)] $h \ge 0$. In this case a solution does not exist because $\left( e^{-{h}/{T}}-1\right)$ is always negative or zero.
\item[ii)] $h < 0$. This range of the chemical potential is allowed. So we can proceed with the same iterative proof as in the case of $NLS$ to establish the existence and uniqueness of the solution of (\ref{tba}).
\end{itemize}
Here we have one of the most striking differences between $LL$ and $NLS$ models: while in the latter, all the values of $h$ are allowed (although the thermodynamics, and especially $T \rightarrow 0$ limit, drastically depend on its sign), in the former, only negative values of the chemical potential are allowed.

Before we move to discuss the case of zero temperature, $T=0$, it is instructive to compare our situation with $NLS$ when parameter $c$ (the strength of the delta-functional potential) is going to zero. This is relevant, because all the specifics of our model is due to effective {\it turning-off} of the interaction when one of the momenta is approaching zero. In the case of the $NLS$ model, the $TBA$ equation takes exactly the same form (\ref{tba}), where the kernel now is given by
\begin{equation}
 {K}_c(\mu , p )=\frac{2c}{c^2 + (p-\mu)^2}
\label{kernls}
\end{equation}
Once again, the naive setting $c$ to zero would lead to a theory of free bosons. But it is easy to see that this is not the case for $c\longrightarrow +0$, because of the property of ${K}_c(\mu , p )$ analogous to (\ref{kersing})
\begin{equation}
 \lim_{c\rightarrow + 0}{K}_c(\mu , p )=2\pi \delta (\mu - p)
\label{kernlssing}
\end{equation}
which allows to solve $TBA$ exactly with the solution being
\begin{equation}
\varepsilon(p)= T \ln \left( e^{{(p^2-h)}/{T}}-1\right)
\label{nlssolution}
\end{equation}
(Notice the striking similarity between (\ref{epsilon}) and (\ref{nlssolution})!) The analysis of this equations also demonstrates clear difference of the system for $h \ge 0$ and $h < 0$.

Now we give a brief consideration of the zero temperature limit for the case of the allowed chemical potential, $h\in (-\infty, 0)$. From (\ref{epsilon}) one can immediately see that in the zero temperature limit $$\varepsilon (0) = -h >0$$
Using the standard iterative proof of Yang and Yang (see comment after (\ref{epsilon})), we easily establish that this is the minimum of the function $\varepsilon (p)$. As a consequence, $\varepsilon (p)$ is always positive for $T=0$.\footnote{Let us remind, that in the $su(2)$ case the perturbative calculations \cite{Klose:2006dd} lead to instability. The main assumption there is the quantum integrability, which was verified in the first non-trivial order in \cite{Melikyan:2008cy}.} Then, from (\ref{tba}) we find that in this limit we have
\begin{equation}
\varepsilon (p) = p^2 - h > 0
\label{disper}
\end{equation}
In fact, Eq. (\ref{disper}) is exactly what one has in the case of $NLS$ model for $h<0$ and $T=0$, so all the results of the $NLS$ model are applicable here. The main difference is that the other sector of $h$, namely, the $h\in [0, \infty) $ sector is missing in the $su(1,1)$ $LL$ model.

To construct the excitations, we should again start from the general $T\neq 0$ case. Using the continuous form of (\ref{be2}) one can write the analogue of the Lieb integral equation:

\begin{equation}
1+\int \limits_{-\infty}^{\infty}\widetilde{K}(p,\lambda )\rho_p (\lambda
)d\lambda =2\pi \rho_t (p)  \label{lieb3}
\end{equation}
where, as usual, $\rho_p (p)$ and $\rho_t (p)$ are the distribution densities respectively of particles and vacancies.

The excitations are obtained from (\ref{be2}) in the standard manner, we
refer the reader to the monograph \cite{Korepin:1997bk} for complete details of the similar derivation for the $NLS$ model. For
example, a one-particle excitation with the momentum $\lambda_{p}$ is described by the
equation:
\begin{equation}
-\Theta (1/p-1/\lambda_{p})+2\pi F(p| \lambda_{p})-\int \limits_{-\infty}^{\infty} \widetilde{K}
(\mu , p ) \frac{F(\mu | \lambda_{p})}{1+e^{\varepsilon(\mu)/T}}d\mu =0  \label{exc1}
\end{equation}
where the shift function $F(p| \lambda_{p})\equiv \frac{p - \widetilde{p}}{\delta p}$ describes transition from the set of initial momenta $\{ p \}$ to the set of excited momenta $\{ \widetilde{p} \}$. Similarly,
one can easily write the equations describing the hole excitation as well.
Using (\ref{exc1}), the excitation energy takes the form\footnote{It can be seen from (\ref{exc1}) that the function $F(p| \lambda_{p})$ has a discontinuity at $p=0$. However, it is easy to see that the integral (\ref{exc2}) is well-defined.}
\begin{equation}
\Delta E(\lambda_p)=\varepsilon _{0}(\lambda_p)-\int \limits_{-\infty}^{\infty}\varepsilon
_{0}^{\prime }(\mu ) \frac{F(\mu | \lambda_{p})}{1+e^{\varepsilon(\mu)/T}} d\mu  \label{exc2}
\end{equation}
where the energy of the free particle $\varepsilon _{0}(p)=p^{2}-h,$. It can be shown that $\Delta E(p)=\varepsilon (p),$ where the latter function is
determined from the $TBA$ integral equation (\ref{tba}). As usual, this demonstrates that $\varepsilon (p)$ is indeed the particle excitation energy.
The proof consists of manipulating with TBA equation (\ref{tba}) and equation (\ref{exc1}). In particular, using the equation (\ref{disper}), it is easily seen that $\Delta E(p)$ and $\varepsilon (p)$ have the same $p^{2}-h$
asymptotic behavior in the $T\longrightarrow 0$ limit.

Finally, we only mention here that the computation of the correlation functions
at the finite temperature $\left\langle \Omega _{T}\right\vert
S^{+}(x)S(y)\left\vert \Omega _{T}\right\rangle $ requires a separate
investigation, as the unusual Hilbert space structure (\ref{vec1}) and the
scalar product (\ref{scal.pro}) will essentially modify the results. We
leave this to a future investigation.

\section{Conclusion}

We have presented the essential steps to construct the thermodynamics for the
isotropic $su(1,1)$ $LL$ model, which is an interesting toy-model due to its
singular potential structure. We only used our method of constructing the
self-adjoint extensions and regularized Hamiltonian to derive the Bethe
Ansatz equations, as well as the $S$-matrix factorization property. This
enables us to derive the spectrum and proceed to derivation of the
thermodynamics. Several key features make the thermodynamics of the $LL$ model an interesting system. Firstly, we showed, using non-symmetric and singular behavior of the kernel, that the self-consistency of the $TBA$ equations restrict the chemical potential to be a negative function. This leads to a rather restrictive character of the spectrum. We also showed that the $su(1,1)$ theory is stable, unlike its $su(2)$ counterpart. Another essential conclusion was that one has to start from the $T \neq 0$ and only then consider a careful zero-temperature limit.

One interesting question, which we mention just briefly, is to to compare the finite size corrections for $T=0$ for the energy
with the corresponding corrections at infinite length and finite temperature. Let us remind, that, e.g., in the $NLS$ model it
can be shown than in the first case the finite size corrections have the
form:
\begin{equation}
E=R\int d\mu \rho _{p}(\mu )\left( \mu ^{2}-h\right) +\frac{const}{R}
v_{F}+...  \label{cor1}
\end{equation}
where $R=1/T$, and $v_{F}$ is the Fermi velocity. This is to be compared with $E^{\prime }\equiv TF(T)/L$ , at
the limit $L\longrightarrow \infty .$ In this case the expansion has the
form:
\begin{equation}
E^{\prime }=R\int d\mu \rho _{p}(\mu )\left( \mu ^{2}-h\right) +\frac{const}{
R}C_{V}+...  \label{cor2}
\end{equation}
where in the second term $C_{V}$ is the heat capacity at constant volume.
The reason these two expressions do not coincide is, of course, the lack of
Lorentz invariance in the $NLS$ model. However, these computations can be
easily repeated for the relativistic $AAF$ model, and we expect these two
expressions to coincide \cite{Zamolodchikov:1989cf}. On the other hand, in the case of the $LL$ model we encounter a problem, since the Fermi point, and therefore, the Fermi velocity are not defined. In this case, one has to consider the finite size correction at \emph{finite} temperature and then consider the $L\longrightarrow \infty$ limit. We leave this to a future investigation.

We again emphasize that the $LL$ model should be considered as a toy model to understand the nuances of the singular potentials. It will be interesting, as a next step, to apply the results of this paper to the $AAF$ model \cite{Alday:2005jm}, which shares many general features with the $LL$ model, but which is more interesting due to its relativistic invariance, as well as non-trivial bound states structure. Other interesting and related developments of our approach include generalization of our method to the anisotropic $LL$ model and, what should be more challenging due to problems mentioned above, to the $su(2)$ $LL$ model. All these problems are currently under investigation, and we hope to report on the results in the future.

\section*{Acknowledgment}
We would like to thank The National Council for Scientific and Technological Development (CNPq), Brazilian Ministry of Science and Technology (MCT), and Instituto Brasileiro de Energia e Materiais (IBEM) for the partial support of our research.

\bibliographystyle{utphys}
\bibliography{ll_therm_final}
\end{document}